\documentclass[twocolumn,showpacs,superscriptaddress,prb,amsmath,amssymb,letterpaper,floatfix]{revtex4}
\usepackage{times}
\usepackage{amsmath,bm,amsfonts,dsfont}
\usepackage{graphicx}
\graphicspath{./figures/}
\usepackage{latexsym}

\renewcommand{\v}[1]{\ensuremath{\mathbf{#1}}} % for vectors
\newcommand{\gv}[1]{\ensuremath{\mbox{\boldmath$ #1 $}}}
\newcommand{\avg}[1]{\left< #1 \right>} % for average
\newcommand{\pd}[2]{\frac{\partial #1}{\partial #2}}
\newcommand{\ket}[1]{\left| #1 \right>} % for Dirac bras
\let\baraccent=\= % rename builtin command \= to \baraccent
\renewcommand{\=}[1]{\stackrel{#1}{=}} % for putting numbers above =
\newcommand{\bk}{\v{k}}

\newcommand{\bq}{\v{q}}

\newcommand{\tm}{\tilde{m}}

\newcommand{\HM}{\mathcal{H}}

\newcommand{\Jeff}{J_{\textrm{eff}}}
\newcommand{\Jin}{J_{\textrm{in}}}
\newcommand{\Jout}{J_{\textrm{out}}}

\newcommand{\vphi}{\gv{\phi}}

\newcommand{\eref}[1]{(\ref{#1})}

\begin{document}

\title{Interplay between Lattice Distortion and Spin-Orbit Coupling in
Double Perovskites}

\author{Tyler \surname{Dodds}}
\affiliation{Department of Physics, University of Toronto,
Toronto, Ontario M5S 1A7, Canada}

\author{Ting-Pong \surname{Choy}}
\affiliation{Department of Physics, University of Toronto,
Toronto, Ontario M5S 1A7, Canada}
\affiliation{Instituut-Lorentz, Universiteit Leiden, P.O. Box 9506,
2300 RA Leiden, The Netherlands}

\author{Yong Baek \surname{Kim}}
\affiliation{Department of Physics, University of Toronto,
Toronto, Ontario M5S 1A7, Canada}
\affiliation{School of Physics,
Korea Institute for Advanced Study, Seoul 130-722, Korea}

\date{\today}

\begin{abstract}

We develop anisotropic pseudo-spin antiferromagnetic Heisenberg models for
monoclinically distorted double perovskites.
We focus on these A$_2$BB$'$O$_6$ materials
that have magnetic moments on the $4d$ or $5d$ transition metal B$'$ ions, which
form a face-centered cubic lattice.
In these models, we consider local $z$-axis distortion of B$'$-O octahedra,
affecting relative occupancy of $t_{2g}$ orbitals,
along with geometric effects of the monoclinic distortion and spin-orbit coupling.
The resulting pseudo-spin-$1/2$ models are solved in the saddle-point
limit of the Sp($N$) generalization of the Heisenberg model.
The spin $S$ in the SU(2) case generalizes as a
parameter $\kappa$ controlling quantum fluctuation in the Sp($N$) case.
We consider two different models that may be appropriate for these systems.
In particular, using Heisenberg exchange parameters for La$_2$LiMoO$_6$ from a
spin-dimer calculation, we conclude that this pseudo-spin-$1/2$ system may order,
but will be very close to a disordered
spin liquid state.

\end{abstract}

\pacs{75.10.Jm, 75.10.Kt}

\maketitle

\section{Introduction}

Geometrically frustrated magnets have been of great
recent interest, and are a common starting point in search of exotic
ground states.\cite{moessnerfrustration,greedanfrustration}  
One class of such frustrated antiferromagnets 
is found in the double perovskite oxides, which host a wide 
range of interesting behavior.\cite{ericksonBaNaOsO,krockenbergermogare,
battleBaYRuO,kobayashiSrFeMoO,rameshmultiferroic,katoreperovskites}
These compounds of chemical formula A$_2$BB$'$O$_6$ feature ordered,
interpenetrating face-centered cubic (FCC) lattices of the B and B$'$ ions
when the charge difference between these ions is large.\cite{andersondperov}
Both B and B$'$ transition metal ions are octahedrally coordinated by oxygen.
A geometrically frustrated FCC lattice is obtained when only the B$'$ ions
are magnetic.

A conventional picture of isotropic antiferromagnetic superexchange is
insufficient for these materials. Altering this picture are
two important effects considered in our work.  The first effect is spin-orbit
coupling, which is relevant for the $4d$ and $5d$ transition metal ions that
comprise the magnetic sites.  
Spin-orbit coupling has been seen to lead to increased correlation effects,
particularly in materials containing $5d$ Ir ions.
This is responsible for 
topological insulating behavior,\cite{shitadekatsura} 
particularly in the pyrochlore iridates,
\cite{pesinbalents,matsuhirawakeshima,fukazawamaeno,nakatsujiprir,wanturner,
yangkim}
the Mott insulator ground state of Sr$_2$IrO$_4$,
\cite{cavabatlogg,kimjinexpt,kimscience,jinldasriro,moonjin,jackeli,
wangsenthil}
and the potential spin-liquid ground state of Na$_4$Ir$_3$O$_8$
\cite{zhouleenairo,okamatonairosl,balentsnairo,lawlerkeekimnairo,normanmicklitz,
hopkinsonnairo,lawlerparamekanti,kimpodolsky,podolskyslmetal}
and honeycomb compounds A$_2$IrO$_3$.\cite{chaloupka}
Octahedral crystal fields favor the $t_{2g}$ \textit{d}-orbitals, which
have an effective orbital angular momentum $L_{\textrm{eff}}=1$, up to a sign
difference.  Combined with $S=1/2$ spin angular momentum, the pseudo-total
angular momentum states of $\Jeff = 1/2$ and $\Jeff = 3/2$ result.
In this case, the quadruplet of $\Jeff=3/2$ states 
form a lower energy manifold than the other two states of
$\Jeff=1/2$.\cite{chensoc} 
The second effect is geometrical distortion from the cubic case; monoclinic
distortion is commonly seen in double perovskites.\cite{andersondperov} 
Lowered symmetry from the monoclinic distortion will spoil the exchange 
isotropy directly, and introduce new exchange pathways.
One particularly important result is
the local $z$-axis compression or expansion of the B$'$-O octahedra, which
we refer to as a \textit{tetragonal distortion} of these octahedra.  While
the octahedral crystal field favors the $t_{2g}$ orbitals over the $e_g$ ones,
the tetragonal distortion will split the $t_{2g}$ levels.  In the case of a
local $z$-axis compression, the $d_{xy}$ orbital is favored
to be occupied, while an expansion favors the
$d_{xz}$ and $d_{yz}$.
All of these effects will generate the
anisotropic interactions that form the focus of our models.

\begin{figure}[b!]
        \includegraphics[scale = 0.35]{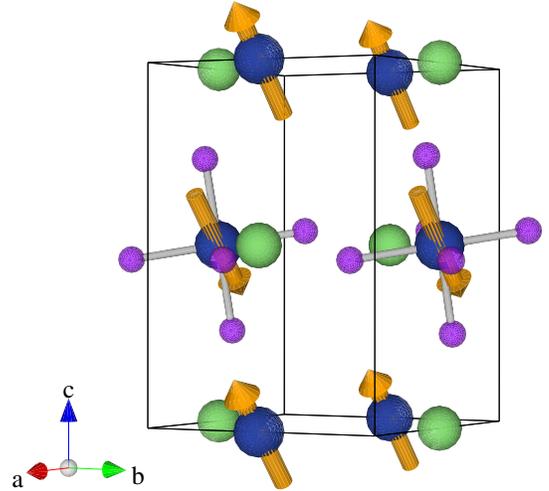}
        \caption{(color online). Magnetic ordering 
	(Type I antiferromagnetic) of the spin-$3/2$
        Ru in La$_2$LiRuO$_6$ (blue, with arrows).\cite{battleLaLiRuO}
        Also shown are the non-magnetic Li (light green) atoms, and
        two of the Ru-O (purple) octahedra, showing the effects
	of monoclinic distortion.}     
        \label{fig:laliruo}
\end{figure}

The role of spin-orbit coupling in the undistorted cubic double
perovskites has been carefully considered by Chen \textit{et al.}
for materials of $d^1$ electronic configuration.\cite{chensoc}
In this work, we focus on the $4d^1$ and $5d^1$ monoclinically distorted 
double perovskites, and consider the
quantum pseudo-spin-$1/2$ models that result, as explained
in the main body of the paper. 
We are particularly interested in the case of 
a local B$'$-O $z$-axis compression, where orbital degeneracy is absent.
La$_2$LiMoO$_6$ is a candidate for such a material, while the otherwise
isostructural Sr$_2$CaReO$_6$ features instead a $z$-axis
expansion of the octahedra.
La$_2$LiMoO$_6$ shows no magnetic ordering down to 2 K from either
heat capacity or neutron diffraction; however, $\mu$SR measurements show
evidence of short-range correlations developing below 20 K.\cite{aharendperovMo}
The Curie-Weiss temperature is negative, $\theta_C = -45$ K, indicating
predominant antiferromagnetic superexchange.
In contrast, Sr$_2$CaReO$_6$ shows spin-freezing behavior 
below 14 K.\cite{wiebeSrCaReO}

In the present work, we use the Sp($N$) generalization of Heisenberg models
to describe these systems.\cite{sachdevkagome,readsachdev,sachdevread}
This generalization provides a unifying framework to
study the effect of spin magnitude, from semiclassical
ordering at ``large spin'' to possible spin liquid phases for
``small spin''.

The ability to capture ``large-spin'' magnetic order may help to 
describe the higher-spin analogues of $d^1$ double perovskites.
In particular, the ``spin-$3/2$'' analogue of La$_2$LiMoO$_6$ is the isostructural
La$_2$LiRuO$_6$, whose $4d^3$ configuration occupies all three $t_{2g}$
orbitals. 
Since the effective magnetic moment is close to the 
spin-$3/2$-only moment, there is only slight renormalization
due to spin-orbit coupling, and 
intra-orbital Coulomb repulsion
is the dominant effect in determining orbital occupancy.
We model this material with a spin-$3/2$
Heisenberg model, given the lack of orbital degeneracy, providing
a test for Sp($N$)-predicted ordering at spin larger than $1/2$.
In fact, La$_2$LiRuO$_6$ shows type I antiferromagnetic ordering 
below 30 K,\cite{battleLaLiRuO} 
where spins are aligned on each $x$-$y$ plane but antiparallel on
the $x$-$z$ and $y$-$z$ planes, as seen in Figure \ref{fig:laliruo}.
This is consistent with the results in the semi-classical (``large spin'') limit
of our Sp($N$) model.
In contrast, an appropriate pseudo-spin-$1/2$ anisotropic Heisenberg model
for La$_2$LiMoO$_6$ leads to the conclusion that this system must be very close
to a spin liquid state. This may be consistent with the absence of magnetic
order down to 2 K seen in experiment.\cite{aharendperovMo}

The rest of the paper is organized as follows.
In \S \ref{sec:model}, we discuss the effects of monoclinic
distortion and spin-orbit coupling. This leads us to consider
two different models, the \textit{planar anisotropy} and 
\textit{general anisotropy} models, 
each taking the form of a pseudo-spin
Heisenberg model.
In \S \ref{sec:classicalresults}, we solve for the
classical spin ordering of both of these models.
In \S \ref{sec:meanfieldtreatment}, we describe the Sp($N$) generalization
of the Heisenberg model and its mean-field treatment.
Results of this mean-field treatment are shown in \S
\ref{sec:planaranisotropy} for the planar anisotropy model,
and in \S \ref{sec:generalanisotropy} for the general anisotropy model.
An extension to finite temperature is discussed in \S
\ref{sec:finitetemperature}.
In \S \ref{sec:discussion}, we summarize our results and
discuss extensions of this work.

\section{Model}
\label{sec:model}

In modelling monoclinically distorted double perovskites with $4d$ or $5d$ 
magnetic ions, there are two important effects of the monoclinic distortion 
that should be considered in conjunction with spin-orbit coupling.
The first effect of monoclinic distortion is
local $z$-axis compression or expansion of the B$'$-O octahedra,
which affects orbital occupation. The second is the change of
orbital orientation due to the geometric distortion, which
affects overlap integrals and the resultant interactions.
We will derive our models by considering the effect of distortion
and spin-orbit coupling on the interactions between $t_{2g}$ orbitals.

One motivation for our models comes from
a spin-$1/2$ Heisenberg model obtained
via spin-dimer calculation for the isostructural 
monoclinically distorted double perovskites La$_2$LiMoO$_6$ and
Sr$_2$CaReO$_6$.\cite{aharendperovMo} 
In this method, the tetragonal compression (or expansion) of these materials
was modelled by assuming occupation of only the $d_{xy}$ orbitals (or equal
occupation of only the $d_{xz}$ and $d_{yz}$ orbitals).
This method is also sensitive to the effect of the geometric changes
resulting from the distortion.
However, spin-orbit coupling was not considered, so that the 
assumed orbital occupation will be slightly incorrect.
The result is an anisotropic $S=1/2$ Heisenberg model, with estimates for the
relative strengths of the couplings, seen in Table \ref{tab:jparams}.

\subsection{Interactions}

\begin{figure}[htp]
	\includegraphics[scale = 0.25]{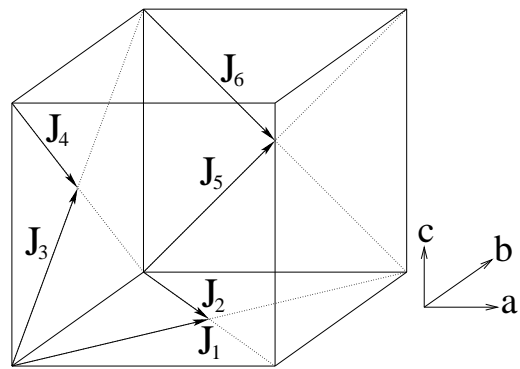}
	\caption{Nearest-neighbor lattice vectors $\gv{\delta}_n$ and
	associated nearest-neighbor couplings $J_n$ for the FCC lattice.}
	\label{fig:fcclattice}
\end{figure}

To understand the effects of the monoclinic distortion
and spin-orbit coupling, we first look at the
interactions between neighboring $t_{2g}$ orbitals in the case of cubic symmetry, 
as have been considered in detail by Chen {\textit{et al}}.\cite{chensoc}
To facilitate this,
we show the six nearest-neighbor directions
$\gv{\delta}_n$ for the FCC lattice 
in Figure \ref{fig:fcclattice}. 
Without distortion, the $a$, $b$ and
$c$-axes are simply the Cartesian $x$, $y$ and $z$-axes.
The strongest interaction is antiferromagnetic superexchange, involving 
sites and orbitals lying in the same plane. For instance, $d_{xy}$ orbitals
on neighboring sites along the $x$-$y$ plane will interact antiferromagnetically.
Ferromagnetic interactions between sites on a plane will couple orbitals lying
on that plane to orbitals lying perpendicular to it.\cite{chensoc}
Along the $x$-$y$ plane,
$d_{xy}$ orbitals interact ferromagnetically with neighboring $d_{yz}$ and
$d_{xz}$ orbitals.
Quadrupole-quadrupole interactions also exist between all $t_{2g}$ orbitals
on neighboring sites, due to different orientations of the quadrupole moments
of these orbitals.

\subsection{Monoclinic Distortion}

The first effect of monoclinic distortion 
is the local $z$-axis distortion of the $B'$-O octahedra, a
compression for La$_2$LiMoO$_6$ and an expansion for Sr$_2$CaReO$_6$.
This splits the degeneracy of the three $t_{2g}$ orbitals.
The $d_{xz}$ and $d_{yz}$ orbitals will remain degenerate, but the
$d_{xy}$ orbital will have a lower energy for a compression and a higher
energy for an expansion.
Consequently, the occupation of the $d_{xy}$ orbital will be
favored or disfavored compared to occupation of the other two orbitals.
This is taken as a very important effect in the spin-dimer calculation
to explain the relative anisotropies of the two materials.\cite{aharendperovMo}

The second important effect of the monoclinic distortion is 
a global $c$-axis elongation, and rotation of the B$'$-O octahedra, affecting
the overlap of the occupied orbitals, which
are now tilted out of plane. 
An example of this, in the case of La$_2$LiRuO$_6$, is shown
in Figure \ref{fig:laliruo}.
The $d_{xy}$ orbitals, for instance, are
tilted out of the $a$-$b$ plane, and will have some interaction 
with $d_{xy}$ orbitals on neighboring planes.
In this fashion, many new exchange pathways will contribute at the 
nearest-neighbor level.

These effects generate a significant amount of exchange anisotropy in the
spin-dimer calculation.\cite{aharendperovMo}
The relative coupling strengths estimated by spin-dimer calculation for 
La$_2$LiMoO$_6$ and Sr$_2$CaReO$_6$
can be seen in Table \ref{tab:jparams}.
\begin{table}
	\begin{tabular}{c|c|c|c|c|c|c}
	Material & $J_1$ &$J_2$ & $J_3$ & $J_4$ & $J_5$ & $J_6$ \\
	\hline
	La$_2$LiMoO$_6$ & 0.14 & 1.0 & 0.014 & 0.014 & 0.00043 & 0.00043 \\
	Sr$_2$CaReO$_6$ & 0.87 & 1.0 & 0.16 & 0.16 & 0.25 & 0.25 
	\end{tabular}
	\caption{Relative strengths of Heisenberg couplings, given in
	Figure \ref{fig:fcclattice}, from the spin-dimer calculation
	of Aharen \textit{et al}.\cite{aharendperovMo}}
	\label{tab:jparams}
\end{table}
Interactions between $x$-$y$ planes in La$_2$LiMoO$_6$ are relatively
weak, as expected from dominant in-plane $d_{xy}$-$d_{xy}$ antiferromagnetic
interaction and $c$-axis elongation.
We note that further in-plane anisotropy is significant, due to the strong
effect of Mo-O octahedra rotation upon $d_{xy}$ orbital overlap.  In
Sr$_2$CaReO$_6$, intra-plane interactions are still larger than inter-plane
interactions, even though the superexchange between $d_{xy}$
orbitals is not present. 
The only in-plane superexchange processes occur through tilted $d_{xz}$ or
$d_{yz}$ orbitals.  Nevertheless, the
inter-plane interactions are significantly stronger than in La$_2$LiMoO$_6$.
The length of the unit cell along the $c$ axis is significantly larger
than along the $a$ or $b$ axes, which could explain the smaller 
inter-plane coupling compared to the intra-plane one.
For both materials, however, the planar anisotropy of the couplings is clear,
and effects of both geometrical distortion and orbital occupation are important.

\subsection{Spin-Orbit Coupling}

Beyond monoclinic distortion, we now consider spin-orbit
coupling, which can be important in the $4d$ and $5d$ magnetic ions
commonly seen in the double perovskites.
For instance, spin-orbit coupling in octahedrally coordinated Mo$^{5+}$ is
estimated to be on the order of 0.1 eV.\cite{vulfson}  
The effect of spin-orbit coupling on the 
$t_{2g}$ orbitals of octahedrally 
coordinated ions is a well-studied problem. 
When the octahedral crystal field splitting is significantly large
compared to the spin-orbit coupling, we may project out the $e_g$ states.
Upon projection, the $L=2$ orbital angular momentum for the $d$ orbitals
looks like a $L=1$ pseudo-angular momentum operator $\v{l}$ up to
a sign change, where $\v{L} \to -\v{l}$.
This $L_{\textrm{eff}}=1$ pseudo-orbital angular momentum combines with the
$S={1}/{2}$ angular momentum of the single electron to create states of
effective total angular momentum $\Jeff=3/2$ and $1/2$.
The spin-orbit coupling $\lambda \v{L} \cdot \v{S}$
breaks the degeneracy of these states,
where the four $\Jeff=3/2$ states have an energy $3\lambda/2$ lower
than the two $\Jeff=1/2$ ones.
These $\Jeff=3/2$ states are written in terms of the $t_{2g}$ ones
as
\begin{align}
	\ket{\frac{3}{2},\frac{3}{2}} = 
	\frac{1}{\sqrt{2}}\left(
	-\ket{yz,\uparrow} + i\ket{xz,\downarrow}	\right)
	\nonumber \\
	\ket{\frac{3}{2},\frac{1}{2}} = 
	\frac{1}{\sqrt{6}}\left(-\ket{yz,\downarrow} + i\ket{xz,\downarrow}
	+ 2\ket{xy,\uparrow} \right)
	\nonumber \\
	\ket{\frac{3}{2},-\frac{1}{2}} = 
	\frac{1}{\sqrt{6}}\left(\ket{yz,\uparrow} + i\ket{xz,\uparrow}
	+ 2\ket{xy,\downarrow} \right)
	\nonumber \\
	\ket{\frac{3}{2},-\frac{3}{2}} = 
	\frac{1}{\sqrt{2}}\left(
	\ket{yz,\downarrow} + i\ket{xz,\uparrow}	\right).
	\label{eqn:jthreehalfstates}
\end{align}
With a $d^1$ configuration, the occupancy of the $d_{xy}$ orbital upon
projection to these states is given by\cite{chensoc}
\begin{align}
	n_{i,xy} = \frac{3}{4} - \frac{1}{3}(j_i^z)^2.
	\label{eqn:xyorbitaloccupancy}
\end{align}
The occupation operators for the other $t_{2g}$ orbitals are given by
cyclic permutation of the $x,y,z$ indices, and the single-occupancy
constraint $n_{i,xy} + n_{i,xz} + n_{i,yz} = 1$ is satisfied.

The effect of projection onto this $\Jeff=3/2$ subspace, due to large
spin-orbit coupling, has been considered by
Chen {\textit{et al.}} for the cubic materials.\cite{chensoc}
The Hamiltonian can be written in terms of the orbitally-resolved
spin operators, such as $\v{S}_{i,xy} = \v{S}_in_{i,xy}$.
Upon projecting to the $\Jeff = 3/2$ states, these orbitally-resolved
spin operators contain terms both linear and cubic in $\v{j}$.
The resulting Hamiltonian, containing terms of 4$^{\textrm{th}}$
and 6$^{\textrm{th}}$ order in $\v{j}$, leads
to interesting multipolar behavior.\cite{chensoc}

When spin-orbit coupling is much larger than the local $z$-axis crystal field,
the $\Jeff=3/2$ states provide the relevant starting point, rather than the
$t_{2g}$ orbitals. However, one can consider the general splitting of
$t_{2g}$ orbital degeneracy in the presence of both spin-orbit coupling and
the local $z$-axis distortion. We can model each site with a local
Hamiltonian 
$\HM_{\mathrm{loc}} = \Delta \left[ (l^z)^2 -2/3\right] 
- \lambda \v{l}\cdot\v{S}$,
where $\Delta > 0$ is the strength of the crystal field splitting due to local
$z$-axis compression. The case for a local $z$-axis expansion has been
considered by Jackeli and Khaliullin.\cite{jackelikhaliullin} 
We proceed in a similar manner, identifying the relevant low-energy
eigenstates of $\HM_{\mathrm{loc}}$.
Diagonalization of $\HM_{\mathrm{loc}}$ determines
the lowest-energy Kramers pair to be given by
\begin{align}
	\ket{\uparrow}_{G} = 
	\frac{\sin{(\theta)}}{\sqrt{2}}
	\left(i\ket{yz,\downarrow} + \ket{xz,\downarrow}\right)
	- i\cos{(\theta)}\ket{xy,\uparrow},
	\nonumber \\
	\ket{\downarrow}_{G} = 
	\frac{\sin{(\theta)}}{\sqrt{2}}
	\left(-i\ket{yz,\uparrow} + \ket{xz,\uparrow}\right)
	- i\cos{(\theta)}\ket{xy,\downarrow},
	\nonumber \\
	\tan{(2\theta)} = 2\sqrt{2}\lambda/(\lambda+2\Delta).
	\label{eqn:lowedoublet}
\end{align}
The energy difference between the ground and first excited doublets is
given by $-\lambda + (\lambda+2\Delta)(1+1/\cos{(2\theta)})/4$, which goes
to zero as $\Delta\to0$, and approaches $\Delta - {\lambda}/{2}$ 
when $\Delta \gg \lambda$.
We consider the case where this separation is large enough to focus
on the lowest-energy doublet. 
This will require the tetragonal crystal field to be significantly
larger than the exchange coupling $J$, regardless of the relative strength of
spin-orbit coupling.  By projecting out the higher-energy states, we obtain a
pseudo-spin-1/2 model.

Within this projection, we consider the form of the interactions in an
otherwise cubic double perovskite,
beginning with the quadrupole-quadrupole interaction. 
Due to the fixed orbital occupation in \eref{eqn:lowedoublet},
this interaction is constant and will not contribute to our models.
The orbitally off-diagonal ferromagnetic interactions,
of strength $J'$, generate pseudo-spin interactions that
are both spatially and spin-anisotropic.
For our models, we will focus on the antiferromagnetic interactions.
Nearest-neighbor interactions along the undistorted
$x$-$y$, $x$-$z$ and $y$-$z$
planes are given by
\begin{align}
	\HM_{\textrm{AF}} = 
	J\sum_{<ij>\textrm{ in x-y}}
	\left(\v{S}_i\cdot\v{S}_j-\frac{1}{4}\right)n_{i,xy}n_{j,xy}
	\nonumber \\
	+ (xy \to yz) + (xy \to xz),
	\label{eqn:generalantiferromagnetic}
\end{align}
where $n_{i,xy}$ is the occupation operator of the $d_{xy}$ orbital at site
$i$.\cite{chensoc}
Upon projection to the lowest-energy doublet,
we obtain a Heisenberg model in the pseudo-spin-$1/2$
operators $\v{P}_i$,
\begin{align}
	\HM' =
	N\left(-\frac{J}{4} \right) + 
	\sum_{<ij>\textrm{ in x-y}} \cos{(\theta)}^4
	J \v{P}_i\cdot\v{P}_j
	\nonumber \\
	+ \sum_{<ij>\textrm{ in x-z}} 
	\sin{(\theta)}^4\frac{J}{4}\v{P}_i\cdot\v{P}_j
	+ \sum_{<ij>\textrm{ in y-z}} 
	\sin{(\theta)}^4\frac{J}{4}\v{P}_i\cdot\v{P}_j
	.
	\label{eqn:projectedantiferromagnetic}
\end{align}
For $\Delta \ll \lambda$, this result reduces to the one
obtained by Chen \textit{et al.} in the easy-plane limit
of the cubic perovskite model with $J'=0$.\cite{chensoc}
Without an accurate estimate for the strength of Hund's coupling
to Coulomb repulsion,
the ratio $J'/J$ is difficult to ascertain.
However, we note that the easy-plane result of Chen {\textit{et al}.}
is an antiferromagnetic state for $J' < J$.\cite{chensoc}
Consequently, we consider the physical picture of antiferromagnetic
interactions, and 
as a first-order approximation we ignore the ferromagnetic
contributions to the Hamiltonian.

We note that the introduction of spin-orbit coupling results in
a reduction of the magnetic moment compared to the case of
$d_{xy}$ occupation when $\lambda=0$.

\subsection{Planar-Anisotropy and General-Anisotropy Models}

The first, and simpler, of the two models considered in this paper
is concerned primarily with the effects of the tetragonal crystal field splitting.
Without spin-orbit coupling, we see easily from 
\eref{eqn:generalantiferromagnetic} that preferential $d_{xy}$ orbital
occupation leads to anisotropic interactions that are stronger on
the $x$-$y$ planes. In this case, we have a true spin-$1/2$
antiferromagnetic Heisenberg model.
However, considering spin-orbit coupling and tetragonal
distortion leads to the pseudo-spin-$1/2$ antiferromagnetic Heisenberg
model in \eref{eqn:projectedantiferromagnetic}, with a
similar form of anisotropy.
From this, we are motivated to study the pseudo-spin-$1/2$
antiferromagnetic Heisenberg model where coupling along the $x$-$y$ plane
differs from the coupling along the $y$-$z$ and $x$-$z$ planes.
The \textit{planar anisotropy model} is given in terms of 
pseudo-spin-1/2 operators (henceforth referred to as $\v{S}_i$) by
\begin{align}
	\HM_{\textrm{PA}} = 
	\Jin\sum_{<ij>\textrm{ in x-y}}
	\v{S}_i\cdot\v{S}_j
	\nonumber \\
	+ \Jout\sum_{<ij>\textrm{ in y-z}}
	\v{S}_i\cdot\v{S}_j
	+ 
	\Jout\sum_{<ij>\textrm{ in x-z}}
	\v{S}_i\cdot\v{S}_j
	\label{eqn:planaranisotropymodel}.
\end{align}
Both $\Jin$ and $\Jout$ are antiferromagnetic, and 
one can consider this model as a generalization of the
antiferromagnetic model in Eq. \eref{eqn:projectedantiferromagnetic}.
The ratio $\Jout / \Jin$ depends on the strengths of the spin-orbit coupling
and tetragonal distortion of the octahedra, seen in
$\Delta/\lambda$. In addition, it captures
certain geometrical effects of the monoclinic distortion, such as the global
$c$-axis elongation, contributing to the particular planar anisotropy in
\eref{eqn:planaranisotropymodel}.

The other model considered in this paper will include in full
the geometrical effects of the monoclinic distortion.
This will generate many other anisotropic interactions, 
breaking the symmetry of the $x$-$y$ plane. 
Effective pseudo-spin exchange energies will become intrinsically
anisotropic, in addition to the effects of orbital occupation.
We will model these like the spin-dimer calculation does, with
different strengths of the nearest-neighbor couplings shown
in Fig. \ref{fig:fcclattice}. Due to spin-orbit coupling,
the particular parameters $J_n$ in Table \ref{tab:jparams} will not be 
quantitatively correct. Nonetheless, 
we will consider them as a starting point to understand the 
effect of further anisotropy in the interactions. Estimates for corrections
due to spin-orbit coupling are given in 
\S\ref{sec:inplaneoutofplaneanisotropy}.
The \textit{general anisotropy model} is given by
\begin{align}
	\HM_{\textrm{GA}} = 
	\sum_i \sum_n J_n \v{S}(\v{r}_i)\cdot\v{S}({\v{r}_i+\gv{\delta}_n}).
	\label{eqn:generalmodel}
\end{align}

To analyze the model Hamiltonians \eref{eqn:planaranisotropymodel}
and \eref{eqn:generalmodel},
we will use the Sp($N$) generalization of the Heisenberg model, which
offers several advantages. The first is that the parameter
$N$ allows for a controlled expansion, beginning from the
saddle-point solution as $N\to\infty$. The second is that
quantum fluctuations can be controlled by a parameter $\kappa$
(where $\kappa=2S$ in the SU(2) case)
allowing a transition from a classical-spin limit (large $\kappa$)
to one dominated by quantum fluctuations (small $\kappa$).
This may capture a changing value of (pseudo)-spin.
The gapped $Z_2$ spin liquid, obtained as a disordered state in the Sp($N$)
generalization, is often seen as a potential ground state in many 
Heisenberg models.
\cite{wangvishwanath,moessnersondhifradkin}

The Sp($N$) generalization may be capable of naturally capturing the changing
behavior with $S$ seen in the family of magnetic materials
isostructural to La$_2$LiMoO$_6$.
The spin-$3/2$ La$_2$LiRuO$_6$ is magnetically ordered, while
spin-$1/2$ La$_2$LiMoO$_6$ shows short-range
correlations and suppression of magnetic order. 
The isostructural spin-1 La$_2$LiReO$_6$ is more amenable to a multi-orbital model,
and falls outside the scope of these calculations.\cite{aharendperovRe}

\section{Classical Ordering}
\label{sec:classicalresults}

In this section, we solve both planar anisotropy and general anisotropy models
in the limit of classical spins. The magnetic ordering patterns
and wavevectors are determined by the $O$($N$) model, 
where we generalize to $N\to\infty$ components of the spin vector,
as explained in Appendix \ref{sec:onmodel}.
We will see in \S \ref{sec:kappaexpansion} that this corresponds also to the
classical limit of the Sp($N$) model.

\subsection{Planar-Anisotropy Model}

In the planar anisotropy model \eref{eqn:planaranisotropymodel}, 
two phases are found with varying
$\Jout/\Jin$, the ratio of inter-plane to intra-plane interactions.
For $\Jout < \Jin$, the intra-plane interactions create antiferromagnetic
N\'eel order within each $x$-$y$ plane. For $\Jout > \Jin$, the
inter-plane interactions create antiferromagnetic order between planes.

\begin{figure}[htp]
	\includegraphics[width =  6 cm]{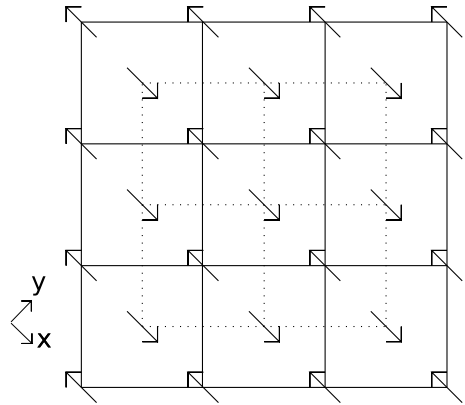}
	\caption{View along the $z$-axis of FCC lattice magnetic ordering of
	the planar anisotropy model for $\Jout > \Jin$.
	The solid lines indicate an $x$-$y$ plane of the FCC lattice, while
	the dotted lines indicate a neighboring plane.
	Spins are aligned on each of the $x$-$y$ planes, but N\'eel ordered
	along $x$-$z$ or $y$-$z$ planes.}
	\label{fig:largekorderingplanar}
\end{figure}

For $\Jout > \Jin$, the ordering wavevector 
$\bq$ is given by
\begin{align} \bq = \frac{\pi}{a/2}\left(0,0,1\right)
	~\mathrm{or}~ \frac{\pi}{a/2}\left(1,1,0\right).
	\label{eqn:interplanewavevec}
\end{align}
Spins on each $x$-$y$ plane are aligned, while
spins on neighboring planes are antiparallel.
N\'eel ordering is found along the $x$-$z$ and $y$-$z$ planes.
The antiferromagnetic interactions between 
$x$-$y$ layers are satisfied, as seen in Figure
\ref{fig:largekorderingplanar}.

\begin{figure}[htp]
	\includegraphics[width =  6 cm]{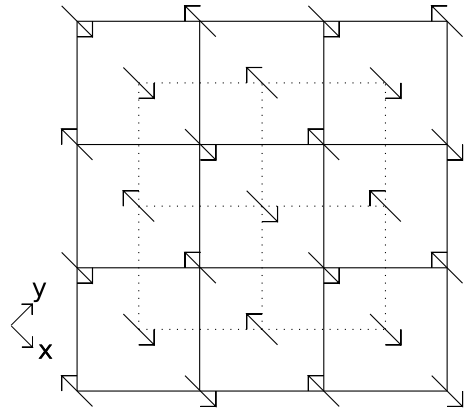}
	\caption{View along the $z$-axis of FCC lattice magnetic ordering of
	the planar anisotropy model for $\Jout < \Jin$ with $k_z=0$.
	The solid lines indicate an $x$-$y$ plane of the FCC lattice, while
	the dotted line indicates a neighboring plane.
	There is N\'eel ordering along each of the
	$x$-$y$ and $y$-$z$ planes, but ferromagnetic ordering along the $x$-$z$
	plane. Also possible is a state where the ferromagnetic ordering is
	along the $y$-$z$ plane instead.}
	\label{fig:largekorderingneel}
\end{figure}

For $\Jout < \Jin$,
the ordering wavevector $\bq$ is 
\begin{align} \bq = \frac{\pi}{a/2}\left(1,0,k_z\right),
	~ \frac{\pi}{a/2}\left(0,1,k_z\right)
	\label{eqn:onorderingneel}
\end{align}
for arbitrary $k_z$.
Each $x$-$y$ plane takes on the 
N\'eel order for a square lattice. 
The degeneracy in $k_z$ indicates that spins on neighboring planes
may take any relative overall orientation.
An example of this ordering, with $k_z=0$, is given
in Figure \ref{fig:largekorderingneel}.
We will see in \S \ref{sec:kappaexpansion} that this degeneracy is broken by
the introduction of quantum fluctuations, choosing $k_z=0$.

Both of these states show Type I antiferromagnetic ordering on the FCC
lattice, where ordering is antiferromagnetic on two of the 
$x$-$y$, $x$-$z$ or $y$-$z$ planes, and ferromagnetic on the other.

\subsection{General Anisotropy Model}

The two parameter sets in Table \ref{tab:jparams} also yield antiferromagnetic
ordering in the $x$-$y$ plane, similar to the $\Jout < \Jin$ case.
However, the degeneracy of $k_z$ is broken here at the classical level,
where $k_z=0$ for both parameter sets.
Ordering as in Figure \ref{fig:largekorderingneel} results.

\section{Sp($N$) Mean Field Theory}
\label{sec:meanfieldtreatment}

\subsection{Sp($N$) Generalization of the Spin Models}
\label{sec:spngeneralization}

The Sp($N$) method is a large-$N$ generalization of the Schwinger boson
spin representation.\cite{sachdevkagome,readsachdev,sachdevread}
In the physical case $N=1$,
Sp(1) is isomorphic to SU(2), and we have the standard Schwinger
boson representation wherein
$\v{S}_{ia} = \frac{1}{2} b^{\dag}_{i\alpha}(\gv{\sigma}_a)_{\alpha\beta}
b_{i\beta}$ and the boson number per site
$b^{\dag}_{i\alpha}b_{i\alpha} \equiv n_b = 2S$ determines
the spin quantum number. Here, $\alpha, \beta = \uparrow, \downarrow$
label the primitive spin-$1/2$ species that comprise the
full spin angular momentum.
We generalize to $2N$ flavors of bosons, where $\alpha=(m,\sigma)$, 
labelled by $m=1\ldots N$, and $\sigma=\uparrow,\downarrow$, transforming under
the group Sp($N$).\cite{sachdevkagome}
$\kappa=n_b/N$ acts in analogous fashion to $2S$ in the SU(2) case, controlling
the strength of quantum fluctuations.

When generalized to Sp($N$), the Heisenberg Hamiltonian
\eref{eqn:generalmodel}, up to constants involving $n_b$,
is written as
\begin{align} \HM = \frac{-1}{2N}\sum_{i}\sum_n J_n
	(\mathcal{J}^{\alpha\beta}b^{\dag}_{i\alpha}b^{\dag}_{i+\delta_n,\beta})
	(\mathcal{J}_{\gamma\nu}b_i^{\gamma}b_{i+\delta_n}^{\nu}).
	\label{eqn:spnspinspin}
\end{align}
Here, $\mathcal{J}_{\alpha\beta}$ is a $2N\times 2N$ block-diagonal
antisymmetric tensor, given by 
\begin{align}
	\mathcal{J}_{m\sigma,m'\sigma'} =
	\delta_{m,m'} 
	\begin{pmatrix}
	0 & 1 \\ -1 & 0
	\end{pmatrix}.
\end{align}

\subsection{Mean-Field States}
The quartic terms in \eref{eqn:spnspinspin} can be quadratically decoupled by
the mean field \begin{align}
	Q_{in}=\frac{1}{N}\left<\sum_{m}\varepsilon_{\sigma\sigma'}
	b^{\dag}_{im\sigma}b^{\dag}_{i+\delta_n,m\sigma'}\right>.
\end{align}

When the boson dispersion becomes gapless,
we allow for a condensate, 
$b_{i1\sigma} = \sqrt{N}x_{i\sigma}\in\mathbb{C}$, where
$\sigma=\uparrow,\downarrow$, so that $\langle b_{i1\sigma}\rangle$ has a
finite expectation value. This will account
for the appearance of long-range magnetic order.

The projective symmetric group analysis may be used to characterize possible
mean-field ground states; for Sp($N$) this has been applied to many other
Heisenberg models.\cite{wangvishwanath}
Qualitatively different states are distinguished by the value of a flux
quantity for plaquettes of the lattice.
The flux on a plaquette of sites $a\ldots z$ 
is defined by the phase $\Phi$ in\cite{tchernyshyovflux}
\begin{align}
	|\Xi| e^{i\Phi} = \sum_{a\ldots z}
	Q_{ab}(-Q_{bc}^*)\ldots Q_{yz}(-Q_{za}^*).
\end{align}
A nearest-neighbor Heisenberg model will favor the zero-flux
states at small $\kappa$,
particularly for plaquettes of smaller length.\cite{tchernyshyovflux}
On the bipartite cubic lattice, for instance,
a translationally invariant choice of $Q_{ij}$ yields zero flux
on any plaquette.
Since the FCC lattice is frustrated, a translationally invariant
$Q_{ij} = -Q_{ji}$, while giving zero flux on most plaquettes, leaves
$\pi$ flux on a small number of plaquettes.
In particular, assuming all $Q$ to be translationally invariant and positive,
the four-site plaquettes with $\pi$ flux 
have sites on both the $x$-$y$ and $y$-$z$ planes, such as
$\v{i},
\v{i}+\gv{\delta}_1,
\v{i}+\hat{y},
\v{i}+\gv{\delta}_3$,
where $\v{i}$ and $\v{i}\pm\hat{y}$ are joined by the plaquette.
There are eight such plaquettes with $\pi$ flux, of a total of thirty-six
four-site plaquettes involving site $\v{i}$.
This provides motivation to consider translationally
invariant mean-field solutions, which we restrict ourselves
to in this work.

\subsection{Mean-Field Hamiltonian}
After decoupling in the site-independent $Q_n$ fields, the Hamiltonian
\eref{eqn:spnspinspin} becomes 
\begin{align} 
	\HM = \sum_{i,n}J_{n}\left[
	-\frac{Q_{n}}{2}\varepsilon_{\sigma\sigma'}\left(
	\sum_{\tm=2}^Nb_i^{\tm\sigma}b_{i+\delta_n}^{\tm\sigma'} + 
	x_i^{\sigma}x_{i+\delta_n}^{\sigma'}N\right) 
	\right. \nonumber \\ \left.
	+ h.c. + 
	\frac{N}{2}\left|Q_{n}\right|^2\right] + \nonumber \\
	\sum_i\mu_i\left(-n_b+\sum_{\tm=2}^Nb^{\dag}_{i\tm\sigma}b_i^{\tm\sigma}
	+N{x}^*_{i\sigma}x_i^{\sigma}\right).
	\label{eqn:decoupledhamiltonian}
\end{align}
Here, the boson number constraint is enforced on average by the inclusion
of the Lagrange multiplier $\mu_i$. We assume tranlsational invariance,
with $\mu_i = \mu$. We have allowed the $m=1$ component to condense,
represented by $x_i^{\sigma} \in \mathds{C}$.

The saddle-point Hamiltonian (for $N\to\infty$) is derived in full in Appendix 
\ref{sec:saddlepointsolution}. The first step is a Fourier transform
defined by $b_i = \frac{1}{\sqrt{N_s}}\sum_kb_ke^{-ik\cdot r_i}$.  
The second step is a Bogoliubov transformation diagonalizing the
Hamiltonian, yielding a quasiparticle energy 
$\omega_k = \sqrt{\mu^2 - (\sum_nJ_nQ_n\sin{(k\cdot\delta_n)})^2}$.
The transformation is defined by $\v{b} = T^{-1}\gv{\gamma}$, where the
Hamiltonian is diagonal in the $\gv{\gamma}$ basis.
The condensate enters only via the total density 
$n=\sum_{k\sigma} |x_{\bk}^{\sigma}|^2$,
and $\pm\bk_1$, the wavevectors of the boson dispersion minimum
where the condensate forms.

We then write the diagonalized Hamiltonian as
\begin{align}
	\frac{\HM}{N_sN} = \sum_{\delta}\frac{J_{\delta}}{2}|Q_{\delta}|^2
	+ \mu\left(-1-\kappa+n
	\right) 
	\nonumber \\ 
	+ n\sum_{\delta}J_{\delta}Q_{\delta}\sin{(k_1\cdot\delta)}
	\nonumber \\ + 
	\frac{1}{N_s}\sum_k\omega_k\left(
	1+\gamma^{\dag}_{k\uparrow}\gamma_{k\uparrow}
	+\gamma^{\dag}_{k\downarrow}\gamma_{k\downarrow}\right).
	\label{eqn:finalmfhamiltonian}
\end{align}

\subsection{Semiclassical Large-$\kappa$ Limit}

\label{sec:kappaexpansion}

We take advantage of the Sp($N$) fluctuation parameter $\kappa$ to look at the
semiclassical magnetic order from the $\kappa\to\infty$ limit.
This provides a link from the classical order of
\S \ref{sec:classicalresults} to the
magnetic order seen at finite $\kappa$.

We begin by approximating the Hamiltonian for $\kappa\gg 1$.
Here, leading-order behavior in the Hamiltonian is of O($\kappa^2$).
Corrections, of O($\kappa$), act to split
degeneracy of the classical ordering.\cite{sachdevkagome}
We have that $Q$, $\mu$ and $n$ are all O($\kappa$) as $\kappa\gg 1$.
$E_C$, the largest contribution to the energy is of O($\kappa^2$):
\begin{align}
	\frac{E_C}{N_sN} = \sum_{\delta}\frac{J_{\delta}}{2}|Q_{\delta}|^2
	+ \mu\left(-\kappa+n\right) 
	\nonumber \\ + 
	n\sum_{\delta}J_{\delta}Q_{\delta}\sin{(k_1\cdot\delta)},
	\label{eqn:largekclassical}
\end{align}
while the first-order quantum correction $E_1$, of O($\kappa$), is given by 
\begin{align}
	\label{eqn:quantumcorrE1}
	\frac{E_1}{N_sN} = -\mu+ \frac{1}{N_s}\sum_k\omega_k,
\end{align}
where $Q$, $\mu$ and $n$ are given by solutions
minimizing the classical energy \eref{eqn:largekclassical}.\cite{sachdevkagome}
The mean-field equations for $E_C$ are easily solved, yielding
$n = \kappa$, 
$\mu= - \sum_{n} J_{n}Q_{n}\sin{(k_1\cdot\delta_n)}$,
and $Q_{m} = -\kappa\sin{(k_1\cdot\delta_m)}$.
We can then write $E_C$ as a function of the minimum wavevector $\v k_1$:
\begin{align}
	\frac{E_C}{N_sN} = -\kappa^2 \sum_{n} \frac{J_{n}}{2}
	\sin^2{(\v{k}_1\cdot\gv{\delta}_n)}.
	\label{eqn:onenergyfinal}
\end{align}
With the boson dispersion minimum at $\pm\v{k}_1$, spin
ordering occurs at the wavevectors $\bq = \pm2\v{k}_1$. 
The minimum of $E_C$ corresponds to an ordering pattern 
equivalent to that of the classical $O$($N$) model (see Appendix
\ref{sec:onmodel} for details).\cite{garanincanalspyrochlore} 
The correction \eref{eqn:quantumcorrE1} can then easily be computed for all
$\v{k}_1$ (with corresponding $Q$, $\mu$, $n$) in the degenerate set of
minima of \eref{eqn:onenergyfinal}.

\section{Planar Anisotropy Model Results}
\label{sec:planaranisotropy}

In this section we study the planar anisotropy model  with in-plane coupling 
$\Jin$ ($J_1=J_2$) and out-of-plane coupling $\Jout$ ($J_3=J_4=J_5=J_6$). 
We study the effect
of quantum fluctuations, controlled by $\kappa$, and coupling anisotropy,
controlled by $\Jout/\Jin$. 
In \S \ref{sec:classicalresults}, we saw classical N\'eel ordering 
on each $x$-$y$ plane. 
The first-order quantum correction $E_1$ in
\eref{eqn:quantumcorrE1} breaks the degeneracy.
After this ``order by disorder'',
the ordering wavevectors are
\begin{align} \bq = \frac{\pi}{a/2}\left(1,0,0\right)
	~\mathrm{or}~ \frac{\pi}{a/2}\left(0,1,1\right),
	\nonumber \\
	\bq = \frac{\pi}{a/2}\left(0,1,0\right)
	~\mathrm{or}~ \frac{\pi}{a/2}\left(1,0,1\right).
	\label{eqn:waveveccorrected}
\end{align}
Spins are aligned along either the $x$-$z$ or $y$-$z$ planes.
Ordering along one such direction was seen in Figure
\ref{fig:largekorderingneel}.

\begin{figure}[htp]
	\includegraphics[width =  8 cm]{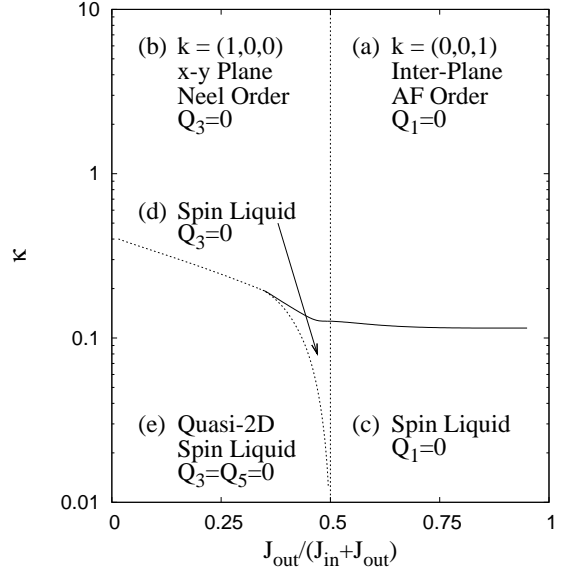}
	\caption{Heuristic phase diagram for the $Q_1,Q_3,Q_5$ ansatz
	of the planar anisotropy model. Note
	that the label $Q_m=0$ indicates that $Q_m$ is negligibly small
	(compared to $\kappa$ and the finite $Q$) in the condensed phase; 
	$Q_m$ is identically zero in the corresponding spin liquid phases.
	Solid lines indicate second-order transitions, while dashed lines
	indicate first-order transitions.}
	\label{fig:disorderedq3phasediag}
\end{figure}

As $\kappa$ is reduced from this limit, we wish to
see the evolution of the ordering wavevector and mean-field parameters.
For small $\kappa$, we investigate the destruction of the ordered
state by quantum fluctuations.
We note that the semiclassical solutions, for all values of 
$\Jout/\Jin$, all feature
$|Q_1|=|Q_2|, |Q_3|=|Q_4|,$ and $|Q_5|=|Q_6|$. 
Motivated additionally by the equality of in-plane couplings,
$J_1=J_2$, and of between-plane couplings,
$J_4=J_4=J_5=J_6$,
we take an ansatz with
$Q_1=Q_2, Q_3=Q_4,$ and $Q_5=Q_6$.
The relative signs, such as between $Q_1$ and $Q_2$,
correspond to making a particular gauge choice.
With such an ansatz, the semiclassical solutions remain
unchanged, with wavevectors
\eref{eqn:interplanewavevec} or
\eref{eqn:waveveccorrected} as appropriate.
Furthermore, relaxing the ansatz suggests that the equivalence 
$|Q_1|=|Q_2|, |Q_3|=|Q_4|,$ and $|Q_5|=|Q_6|$ is retained
down to low $\kappa$.
With this ansatz, we numerically solve the
mean-field equations, given explicitly in Appendix \ref{sec:saddlepointsolution}.
The resulting phase diagram is given in Fig. \ref{fig:disorderedq3phasediag},
in which there are five phases to consider.

\subsection{Inter-plane Antiferromagnetic Order}
\label{sec:interplaneaf}

This state is an extension of the classically ordered state
for $\Jout > \Jin$, with antiparallel magnetization on 
neighboring $x$-$y$ planes. Ferromagnetic ordering is seen
along the $x$-$y$ plane, with N\'eel ordering along the
$x$-$z$ and $y$-$z$ planes.
In this state, the intra-plane $Q_1 = Q_2$ is significantly
smaller than the intra-plane $Q_3$ through $Q_6$.
The ordering wavevector has only small corrections to the classical
result \eref{eqn:interplanewavevec}.

\subsection{$x$-$y$ Plane N\'eel Order}
\label{sec:xyplaneaf}

This state is an extension of the classically ordered state
for $\Jout < \Jin$, with N\'eel order on the $x$-$y$ planes.
It is characterized by large $|Q_1|=|Q_2|$ within the $x$-$y$ plane.
Of the two independent inter-plane $Q$, one is significantly smaller
than the other, depending on the gauge choice of ferromagnetic order
direction (along the $x$-$z$ or $y$-$z$ plane).
The ordering wavevector has only small corrections to the semiclassical result
\eref{eqn:waveveccorrected}.

\subsection{Inter-Plane Spin Liquid}

This state is a disordered analogue of the inter-plane ordered
state (\S \ref{sec:interplaneaf}) for $\Jout > \Jin$.
However, the intra-plane $Q_1=Q_2$ are identically zero in this state.
While the direct intra-plane correlations are consequently zero, the finite
inter-plane $Q$ prevent the lattice from decoupling.
The minimum wavevector, determining short-range order,
still has only small corrections compared to the ordered
minimum \eref{eqn:interplanewavevec}.
The transition into this state from the intra-plane ordered
state, as $\kappa$ is lowered, is second-order.

\subsection{Three-Dimensional Intra-Plane Spin Liquid}

This state is a disordered analogue of the inter-plane ordered
state (\S \ref{sec:xyplaneaf}) for $\Jout < \Jin$.
However, one of the intra-plane $Q$ is now identically zero,
such as $Q_3=Q_4$. 
The other intra-plane $Q$ is nonzero, but still smaller than
the in-plane $Q_1 = Q_2$, preventing the lattice from decoupling.
As before, the minimum wavevector, determining short-range order,
has only small corrections compared to the ordered
minimum \eref{eqn:waveveccorrected}.
The transition into this state from the intra-plane ordered
state, as $\kappa$ is lowered, is second-order.

\subsection{Quasi-Two-Dimensional Spin Liquid}

In this state, all inter-plane $Q$ vanish: 
$Q_3=Q_4=Q_5=Q_6=0$. The system then consists of decoupled
two-dimensional $x$-$y$ planes in this mean-field theory.
The transitions into this state, from either the ordered or disordered
intra-plane states for $\Jout < \Jin$, are weakly first-order.
The minimum (short-range order) wavevector no longer takes the semiclassical
value, instead taking a different value among the classical 
solutions \eref{eqn:onorderingneel}, with $k_z \sim 0.15$.

\subsection{Tricritical Point and Destruction of Order}

We find a tricritical point at $\tilde{J}_{\textrm{out}} =0.58\Jin$
separating the intra-plane spin-liquid phases from the 
$x$-$y$ plane N\'eel ordered phase.
For $\Jin > \Jout > \tilde{J}_{\textrm{out}}$, the ordered state first
enters the three-dimensional spin-liquid state as
$\kappa$ is decreased. A first-order transition to the
two-dimensional spin liquid follows as $\kappa$ decreases
further. The $\kappa$ range of this three-dimensional spin
liquid narrows as $\Jout$ reaches tricritical point,
as seen in Figure \ref{fig:disorderedq3phasediag}.
For $\Jout<\tilde{J}_{\textrm{out}}$, in-plane coupling pushes
the system to decouple.
However, we expect that the $Q=0$ decoupling seen in all three
mean-field spin liquid states is an artifact of the mean-field theory, and that
$1/N$ corrections will restore a small yet non-zero value to these $Q$.

The critical $\kappa$ value of the destruction 
of magnetic ordering, $\kappa_c$, 
is fairly small in this planar anisotropy model. $\kappa_c$ ranges from $0.1$ for
large $\Jout$ to 0.4 for small $\Jout$. In the physical $N=1$ case,
$\kappa=1$ corresponds to the ``most quantum'' limit of $S=1/2$. Our
$N\to\infty$ solution indicates that ordering is likely to occur, 
even though mean-field theory overestimates ordering.  While
$\kappa_c$ will
differ in the exact $N=1$ theory, the 
values of $\kappa_c\sim0.1-0.4$ are too small to account for
the behavior of La$_2$LiMoO$_6$.

\section{General Anisotropy Model Results}
\label{sec:generalanisotropy}

\subsection{Spin Dimer Parameters}
\label{sec:spindimerparameters}

\begin{figure}[htp]
	\includegraphics[scale = 0.37]{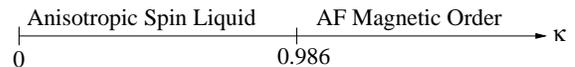}
	\caption{Phase diagram as a function of $\kappa$ for the
	general anisotropy model with parameters for 
	La$_2$LiMoO$_6$, from Table \ref{tab:jparams}.
	For $\kappa$ larger than $\kappa_c=0.986$, the system is
	in a three-dimensional magnetically ordered state,
	as in the semiclassical limit.
	For $\kappa$ smaller than $\kappa_c$, the system is in an
	anisotropic and highly decoupled spin liquid state.
	}
	\label{fig:genanisotropyphasediag}
\end{figure}

We now turn to the particular parameter set in Table \ref{tab:jparams} modelling
La$_2$LiMoO$_6$. 
We saw that the semi-classical limit 
led to Type I antiferromagnetic order, with
N\'eel order on the $x$-$y$ planes. 
As for the planar-anisotropy model, 
we take advantage of coupling symmetry to simplify the mean-field
calculation. We make the ansatz
$Q_3=Q_4$ and $Q_5=Q_6$, since $J_3=J_4$ and $J_5=J_6$.
The semiclassical result satisfies this, while relaxing the ansatz again
suggests this structure carries to low $\kappa$.
Then we numerically solve the resulting mean-field equations.
The mean-field solution finds that 
ordering persists down to $\kappa_c = 0.986$. As in the planar anisotropy case,
the ordering wavevector changes little with $\kappa$, and $Q_5$
remains significantly smaller than the other $Q$.
At $\kappa_c$, there is a weakly first-order phase transition into a disordered
state with $Q_1=Q_3=Q_5=0$. This highly anisotropic 
mean-field solution consists of decoupled
quasi-one-dimensional chains, with $Q_2$ contributing the only non-zero correlation.
The phase diagram for the general anisotropy model with parameters modelling 
La$_2$LiMoO$_6$ is given in
Figure \ref{fig:genanisotropyphasediag}.
As before, we expect $1/N$ corrections to remove this decoupling.

The parameter set for Sr$_2$CaReO$_6$ in Table \ref{tab:jparams} behaves
similarly, although the transition occurs at a smaller $\kappa_c \cong 0.41$,
similar to the values from the planar anisotropy model.

Two comparisons to the planar anisotropy model are relevant. The first is
that at large exchange anisotropy, the mean-field theory continues to 
predict immediate transitions from magnetic order into 
maximally decoupled spin liquid states.
Additionally, this anisotropy stabilizes these decoupled states. For the 
La$_2$LiMoO$_6$ parameters, we see a marked increase in $\kappa_c$,
which falls quite close to 1.
This saddle-point solution suggests that the
$S = 1/2$ system must be very close to the transition
to a spin-liquid state, even if magnetic order eventually appears at very
low temperature. The effect of further quantum or thermal fluctuations
may be sufficient to destroy the order.
This could explain why no long-range order is observed in La$_2$LiMoO$_6$
down to 2 K, while $\mu$SR shows at most short-ranged order.
The distortion of La$_2$LiMoO$_6$ from the cubic perovskite structure is key in
moving beyond the magnetic order predicted by the planar anisotropy model.

\subsection{Corrections to In-Plane and Out-of-Plane Anisotropy}
\label{sec:inplaneoutofplaneanisotropy}

While the Table \ref{tab:jparams} parameters give a good picture of the 
anisotropy of La$_2$LiMoO$_6$, they will not be quantitatively correct.
We wish to look at deviations due to the inclusion of
spin-orbit coupling, from the viewpoint of in-plane and out-of-plane
anistropy. The change in orbital occupation will result in a reduction
of $d_{xy}$-mediated coupling as spin-orbit coupling increases, along with
new contributions, primarily out-of-plane, from 
$d_{xz}$ and $d_{yz}$ occupation.
From these considerations, we estimate changes to $J_n$ so as
to minimize the resulting anisotropy, thus estimating a lower bound
for $\kappa_c$ upon inclusion of spin-orbit coupling.
We determine the effective couplings $J_n$ in a manner similar
to model \eref{eqn:projectedantiferromagnetic}, but with intrinsically
anisotropic exchange modified by orbital occupation.
In general, we have
\begin{align}
	J_n \to \cos{(\theta)}^4 J_n^{xy} + 
	\frac{1}{4} \sin{(\theta)}^4 J_n^{xz,yz},
\end{align}
with $\theta$ as defined in \eref{eqn:lowedoublet}.
While the $\theta = 0$ spin-dimer parameters give $J_n^{xy}$, the
$J_n^{xz,yz}$ are unknown. Since they arise from octahedral tilting,
the in-plane $J_n^{xz,yz}$ will be quite small,
similar to how the out-of-plane $J_n^{xy}$ are small.
Since $0.25\sin{(\theta)}^4$ is also small, we ignore that term by
estimating $J_{1,2}^{xy} = 0$. 
For the out-of-plane interactions, we will
make a large estimate for $J_n^{xz,yz}$ to minimize the out-of-plane
anisotropy, by taking $J_{3,4,5,6}^{xz,yz} = J_2^{xy}$, the largest exchange
scale in the problem.
In terms of the spin-dimer parameters $J_n^{SD}$, we estimate
the change in magnitude of $J_n$ due to the change in orbital occupation
from spin-orbit coupling by taking
\begin{align}
	J_{1,2} = \cos{(\theta)}^4 J_{1,2}^{SD},
	\nonumber \\
	J_{3,4,5,6} = \cos{(\theta)}^4 J_{3,4,5,6}^{SD}
			+ \frac{1}{4} \sin{(\theta)}^4 J_{2}^{SD}.
	\label{eqn:spinorbitrescaling}
\end{align}
For the case of $\lambda \gg \Delta$, we find that $\kappa_c$ reduces
to $0.86$. However, for a moderate case of $\lambda = \Delta$, we find that
there is only a slight reduction in $\kappa_c$ to $0.98$.
For moderate values of $\lambda/\Delta$, these mean-field results indicate
that the system is still close to a disordered state; however,
this will be sensitive to the value of $\lambda/\Delta$.

\begin{figure}[htp]
	\includegraphics[scale = 0.65]{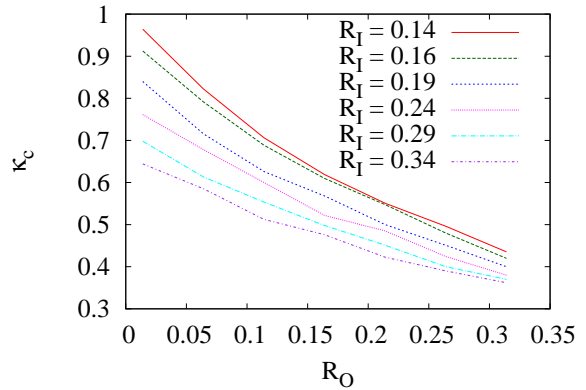}
	\caption{(color online).
	Critical value $\kappa_c$ of destruction of magnetic order.
	Two types of anisotropy are considered. $R_I$ is the ratio of
	the anisotropy within $x$-$y$ planes, while $R_O$ is the ratio
	of anisotropy between these planes.}
	\label{fig:kappacritratio} 
\end{figure}

Exchange anisotropy has shown to be very important, from the results
for the spin-dimer parameters and the spin orbit coupling
rescaled values \eref{eqn:spinorbitrescaling}. To better understand
the combined effect of in-plane and out-of-plane anisotropy, we 
consider a model with slightly less than the full anisotropy,
where $J_1=R_IJ_2$, $J_3=J_4=R_OJ_2$, and $J_5=J_6=R_OR_IJ_2$.
This captures the in-plane ($R_I$) and out-of-plane ($R_O$)
anisotropy, differing from the full anisotropy only
in the very small exchange parameters $J_5$ and $J_6$.
In Figure \ref{fig:kappacritratio} we show $\kappa_c$ as a function of $R_O$,
for several values of $R_I$.
We see that $\kappa_c$ decreases fairly evenly as either $R_O$ or
$R_I$ increases. 
This confirms that both in-plane and out-of-plane anisotropy
are important in securing a large $\kappa_c$.

\section{Finite Temperature}
\label{sec:finitetemperature}

Thermal fluctuations of the quasiparticles in
\eref{eqn:finalmfhamiltonian} introduce, beyond quantum fluctuations,
another mechanism inducing disorder.
At nonzero temperatures, these excitations have a thermal Bose distribution.
The energy $\avg{\HM}$ and the mean-field equations,
\eref{eqn:finalmfhamiltonian} and \eref{eqn:meanfieldeqns},
are modified accordingly.
Thermal fluctuations will reduce magnetic ordering and correlations.
We see different finite temperature behavior depending on the state
(ordered or spin liquid) seen at $T=0$ for a given set of $J_n$ and $\kappa$.

\subsection{Zero-Temperature Disordered Phases}

From disordered phases,
as $T$ increases, the magnitudes of all $Q$ decrease.
The smaller the value of $Q$ at $T=0$, the lower the temperature
at which $Q$ reaches zero. At a large enough temperature,
all $Q$ are zero, describing a perfectly paramagnetic state,
where spins are independent and completely uncorrelated.
This unphysical behavior at high temperature is typical of $N\to\infty$
solutions of Schwinger boson mean-field theories, and disappears
for smaller values of $N$.\cite{tchernyshyovsondhi}

\subsection{Zero-Temperature Magnetic Phases}

From ordered phases,
as $T$ increases, the condensate density $n$ decreases along with the
mean-field parameters $|Q_n|$. It similarly reaches zero at a large enough
$T$.  At large $\kappa$, the transition to the perfect paramagnet state is
first-order, with the system remaining in the ordered state until all $Q$ and $n$
discontinuously jump to zero.  This occurs even for moderate values of
$\kappa$, such as $\kappa \sim 0.5$ in the planar anisotropy model.
For instance, with $\Jout=0.54\Jin$ and $\kappa=0.5$, this transition occurs at
$T=0.44\Jin$. With $\theta_C=-45$ K and $S={1}/{2}$, 
the transition temperature
$T=53$ K, an overestimate to be expected of mean-field theory.

\begin{figure}[htp]
	\includegraphics[scale = 0.65]{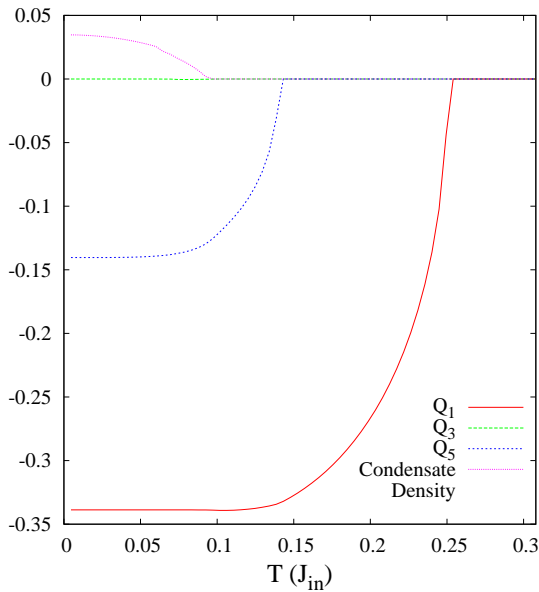}
	\caption{(color online). Mean field and condensate density (magenta)
	destruction with increasing temperature, shown for $\Jout = 0.54\Jin$
	and $\kappa=0.2$ in the planar anisotropy model.
	Above $T \sim 0.1\Jin$, the magnetic ordering is destroyed, leaving
	a thermally disordered state. As $Q_5$ (blue) and $Q_1$ (red) become
	zero, the system enters a two-dimensionally or completely decoupled
	state, respectively.}
	\label{fig:finitetemperature}
\end{figure}

For smaller $\kappa$, close to the disordered
state boundary, the transition is second order. Furthermore, the order can be
destroyed before the $Q$ become zero; the system has a second-order transition to a
thermally disordered state before entering the perfect paramagnet state.
We show such an example in Figure \ref{fig:finitetemperature}.
Here, $\kappa=0.2$, just above the zero-temperature critical $\kappa_c$ for
$\Jout = 0.54 \Jin$. At $T=0$, the transition with varying $\kappa$ went from
ordered state directly into a quasi-two dimensional spin liquid.
At finite temperature, we see that there is a window, $0.1\Jin \lesssim T
\lesssim 0.15\Jin$, where a three-dimensional disordered state exists, in
contrast with the decoupling behavior of the $T=0$ mean-field theory. 

The general anisotropy model with La$_2$LiMoO$_6$ parameters
shows similar behavior. However, at $\kappa=1$, the transition
from the ordered state looks weakly first-order, with the system
directly entering a quasi-two-dimensional decoupled state where
only $Q_1$ and $Q_2$, both in the $x$-$y$ plane, are nonzero.
A fully three-dimensional disordered state is not predicted here by
the finite-temperature mean-field theory. Nonetheless, this case illustrates
how fluctuations destroy magnetic order and inhibit coupling in the
spin-liquid states. As before, we expect $1 / N$ corrections to
further restore correlations.

\subsection{Heat Capacity}

The presence of the perfect paramagnet state 
is an artifact of the mean-field theory.
Regardless, the magnetic contribution to the
heat capacity is an important physical quantity, and can be
reliably calculated in this approach at low temperatures.  $C_V$ is found
straightforwardly from $d{\avg{\HM}}/d{T}$.  In the magnetically ordered
states, we find that $C_V \propto T^3$ at low temperatures. This is expected
from three-dimensional antiferromagnetic spin wave contributions.  In the
disordered states, $C_V \propto \exp{(-\Delta_G/k_BT})$.  $\Delta_G$ scales roughly
with the spin gap, as expected for gapped states.  Unfortunately, the lattice
match material for La$_2$LiMo$_6$ was not useful
in subtracting the lattice contribution to the heat
capacity.\cite{aharendperovMo} 
Without clear data for the magnetic contribution to the specific heat, direct
comparison is not feasible.
For a system close to the ordering transition, such as the general anisotropy
model for La$_2$LiMoO$_6$, the $T^3$ behavior persists only at extremely low
temperatures, further complicating potential comparison.

\section{Conclusion}
\label{sec:discussion}

We have modelled the effects of monoclinic distortion and spin-orbit
coupling in $4d^1$ or $5d^1$ double perovskites. Local $z$-axis distortion of
the magnetic ion-oxygen octahedra changed $d_{xy}$ orbital occupation compared
to the other $t_{2g}$ orbitals.  
Geometrical effects of monoclinic distortion changed orbital overlaps,
introduced multiple exchange pathways and generated significant anisotropy.
Considering spin-orbit coupling in conjunction with the local
$z$-axis crystal field yielded 
a lowest-energy doublet of states
and a pseudo-spin-$1/2$ Heisenberg model from antiferromagnetic interactions.
We considered first the general case where interactions between sites on $x$-$y$
planes differ in strength from interactions between these planes. 
This \textit{planar anisotropy model} was studied for a general
ratio of these two couplings.
Geometrical changes of the monoclinic distortion induce further anisotropy
among the interactions, especially within the $x$-$y$ plane, 
leading to the \textit{general anisotropy model}, studied for particular
parameters modelling La$_2$LiMoO$_6$, estimated from spin-dimer calculation.
\cite{aharendperovMo}
We solved both these models in the saddle-point limit of the
Sp($N$) generalization of the Heisenberg model.
Semi-classical ordering was determined to be Type I antiferromagnetic, with
antiferromagnetic order on two of the $x$-$y$, $x$-$z$, $y$-$z$ planes,
and ferromagnetic order on the other.
The Sp($N$) method connected the semiclassical results to the limit of
large quantum fluctuations.
The large interaction anisotropy of the general anisotropy model
predicted disordering at a relatively large $\kappa_c=0.986$.
The $N=1$ pseudo-spin $1/2$ system was determined to be very close
to a disordered state, even if order sets
in at a low temperature. This could explain the lack of long-range
order seen down to 2 K in La$_2$LiMoO$_6$.
Furthermore, estimates of the effect of spin-orbit coupling on the spin-dimer
calculation parameters of Table \ref{tab:jparams} reduced
$\kappa_c$ only to $0.98$ for moderate strength of spin-orbit coupling.
The system is still close to a disordered state in
this case.

Further experimental and theoretical inquiries follow as natural
extensions of our investigation.
Single-crystal experimental results would be useful, primarily in
determining the short-range ordering wavevector of La$_2$LiMoO$_6$.
Results at temperatures lower than 2 K could determine specifically 
how antiferromagnetic order is being suppressed.
Finally, estimates of the strength of the spin-orbit coupling
and crystal field splitting would guide a more
precise model of the monoclinic distortion.

\begin{acknowledgments}
We are grateful to Leon Balents, John Greedan, Bohm-Jung Yang 
and Jean-Michel Carter for many helpful discussions.
This work was supported by NSERC, the Canada Research Chair program, and the
Canadian Institute for Advanced Research.
We thank the Aspen Center for Physics and the Max Planck Institute for the 
Physics of Complex Systems at Dresden for hospitality, where some parts of this
work were done.  We also acknowledge the Kavli Institute for Theoretical Physics for
hospitality during the workshop ``Disentangling Quantum Many-body Systems",
supported in part by the NSF Grant No. PHY05-51164.
\end{acknowledgments}

\appendix

\section{Classical $O$($N$) Model}
\label{sec:onmodel}

We begin by writing the real-space partition function for the Heisenberg
Hamiltonian on the FCC lattice,
\begin{align}
	Z = \int D\phi D\mu\exp{(-S(\phi,\mu))},~
	\textrm{where }S(\phi,\mu) = 
	\nonumber \\
	\beta \sum_{ij} \left[
	\frac{J_{ij}}{2}\vphi_i\cdot\vphi_j +
	\frac{\mu_i}{2}\delta_{ij}(\vphi_i\cdot\vphi_i - N) \right].
\end{align}
Here, the $O$($N$) model generalizes the spin $\vphi$ from a three-component vector
to an N-component vector.
The first step is to take the Fourier transform defined by
$\vphi_i = \frac{1}{\sqrt{N_s}}\sum_{\v{k}}\vphi_{\v{k}}
	\exp{(-i\v{k}\cdot\v{r}_i)},$ where $N_s$ 
is the number of sites of the lattice.
After the Fourier transform, we have
\begin{align}
	\frac{S}{\beta N_s N} =  - \frac{\mu}{2}
	+ \frac{1}{N_s}\sum_{\bk} 
	|\phi_{\bk}|^2
	\nonumber \\ \times
	\left(\frac{\mu}{2}+ \sum_{n=1}^6 J_{\delta_n}
	\cos{(\bk\cdot\gv{\delta}_n)} \right)
	\\
	Z \propto \int d\mu\prod_{\bk}d\phi_{\bk}d\phi_{\bk}^*
	\exp{(-S)}.
\end{align}
We perform the Gaussian integral over 
$\phi_{\bk}$ and $\phi_{\bk}^*$, giving
\begin{align}
Z \propto \int d\mu
\exp{\left(\frac{\beta\mu}{2} N_sN - \sum_{\bk}
\ln{\left(\frac{D(k,\mu)N\beta}{\pi}\right)}\right)},
\nonumber \\
D(k,\mu) = \frac{\mu}{2} + \sum_{n=1}^6 J_{n}
\cos{(\bk\cdot\gv{\delta}_n)}.
\end{align}
The corresponding saddle-point solution gives $\mu$ from
\begin{align}
	1 = \frac{1}{N_s}\sum_{\bk}\frac{1}{N\beta D(k,\mu)}.
\end{align}
The spin-spin correlation function scales as
\begin{align}
	\left<\gv{\phi}_{\v{k}} \cdot \gv{\phi}_{\v{k}'}\right>
	\propto \delta_{\bk',-\bk} 
	\frac{1}{\beta D(k,\mu)}.
\end{align}
As $\beta\to\infty$, the minimum of $D(\bk,\mu)$ will become the
dominant contribution; magnetic ordering will occur with the
wavevector $\bq$ that minimizes 
$\sum_{n=1}^6 J_{n} \cos{(\bq\cdot\gv{\delta}_n)}$.

\section{Saddle-Point Solution}
\label{sec:saddlepointsolution}
To find the saddle-point solution, we first look at 
the Fourier transform, defined as
$b_i = \frac{1}{\sqrt{N_s}}\sum_kb_ke^{-ik\cdot r_i}$.  
After taking this transform, the Hamiltonian 
\eref{eqn:decoupledhamiltonian} becomes
\begin{align}
	\frac{\HM}{N_sN} = \sum_{n}\frac{J_{\delta_n}}{2}|Q_{\delta_n}|^2
	+ \mu\left(-1-\kappa+\frac{1}{N_s}\sum_kx_{k \sigma}^*x_{k}^{\sigma}
	\right)  
	\nonumber \\ +
	\frac{1}{N_s}\sum_{k n} \left(
	\frac{-J_{\delta_n}Q_{\delta_n}}{2}	\varepsilon_{\sigma\sigma'}
	x_k^{\sigma}x_{-k}^{\sigma'}e^{ik\cdot\delta} + h.c.\right)
	\nonumber \\ + \frac{1}{N_sN}\sum_{mk}
	\begin{pmatrix}b^{\dag}_{km\uparrow}&&b_{-km\downarrow}\end{pmatrix}
		\begin{pmatrix}
		\mu&& B_{k}
		\\ -B_k && \mu
		\end{pmatrix}
	\begin{pmatrix}b_{km\uparrow}\\b^{\dag}_{-km\downarrow}\end{pmatrix};
	\nonumber \\
	B_k = 
	i\sum_{n}J_{\delta_n}Q_{\delta_n}\sin{(k\cdot\delta_n)}
	\label{eqn:hamiltonianft}
\end{align}
where $N_s$ is the number of sites in the system.

The quadratic part of the mean-field Hamiltonian in 
\eref{eqn:hamiltonianft}
is diagonalized by a standard
Bogoliubov transformation.\cite{blaizotripka}
With the quasiparticle energy 
$\omega_k = \sqrt{\mu^2 - (\sum_nJ_nQ_n\sin{(k\cdot\delta_n)})^2}$,
the diagonalized quadratic terms are
\begin{align}
	\frac{1}{N_s}\sum_k\omega_k\left(
	1+\gamma^{\dag}_{k\uparrow}\gamma_{k\uparrow}
	+\gamma^{\dag}_{k\downarrow}\gamma_{k\downarrow}\right).
	\label{eqn:quadraticdiagonalized}
\end{align}
Here, the transformation is defined by $\v{b} = T^{-1}\gv{\gamma}$,
where the columns of $T^{-1}$ are the eigenvectors of ${\eta} M$, 
$M$ is the quadratic Hamiltonian matrix in \eref{eqn:hamiltonianft}, and 
the $2N\times 2N$ ${\eta}$ is given by 
\begin{align*}
	\eta_{\alpha\beta} = \begin{cases}
		\delta_{\alpha\beta} & \alpha \le N \\ 
		-\delta_{\alpha\beta} & \alpha > N
	\end{cases}.
\end{align*}

The structure of the condensate can be determined from the
associated mean-field equation:
$
	\partial {\avg{\HM}} / \partial{x^{\sigma}_k} = 0. 
$ 
The solution to the disordered case ($x=0$) 
has a gapped dispersion. We can track when the
gap vanishes and bosons begin to condense. 
We find that $x_k^{\uparrow}$
is a linear combination of condensates at the minimum wavevectors
$\pm\bk_1$:
$x_k^{\uparrow} = c_1\delta_{k-k_1} + c_2\delta_{k+k_1}$.
We then rewrite the part of the mean-field
energy depending on $x_{\downarrow}$ 
and obtain the mean-field equation
\begin{align}
	0 = \frac{1}{N_sN}\pd{E_{\downarrow}}{x_{k\downarrow}} = 
	\frac{\mu}{N_s}x^*_{k\downarrow} + 
	\nonumber \\
	\frac{1}{N_s}
	\left[\sum_{\delta}J_{\delta}Q_{\delta}\sin{(k\cdot\delta)}\right]
	\left(-c_1\delta_{k,-k_1} -ic_2\delta_{k,k_1}\right).
\end{align}
In the condensed phase, to ensure a gapless dispersion,
$\mu= - \sum_{n}J_{n}Q_{n}\sin{(\bk_1\cdot\gv{\delta}_n)}>0$.
The form of $x_{k}^{\downarrow}$ follows as
$x_{k}^{\downarrow} = -ic_2^*\delta_{k-k_1} + ic_1^*\delta_{k+k_1}$.

We arrive at the diagonalized Hamiltonian \eref{eqn:finalmfhamiltonian}.
From this follow the mean-field equations
\begin{widetext}
\begin{align}
	\frac{1}{N_sN} \pd{E}{\mu} = 0 &=
-1-\kappa+n+\frac{1}{N_s}\sum_k\frac{\mu}{\omega_k},
\nonumber \\
\frac{1}{N_sN}\pd{E}{Q_{m}} = 0 &=
J_{m}Q_{m} + nJ_{\Delta}\sin{(k_1\cdot\delta_m)} -
\frac{1}{N_s}\sum_k
\frac{\sum_{n}J_{n}Q_{n}\sin{(k\cdot\delta_n)}}{\omega_k}
\left(J_{m}\sin{(k\cdot\delta_m)}\right),
\nonumber \\
\frac{1}{N_sN}\pd{E}{n} = 0 &= 
\mu+ \sum_{n}J_{n}Q_{n}\sin{(k_1\cdot\delta_n)}
~\mathrm{(if~}n>0).
\label{eqn:meanfieldeqns}
\end{align}
\end{widetext}

\end{document}